\documentclass[pra,superscriptaddress,twocolumn]{revtex4}
\usepackage{dcolumn}
\usepackage{amssymb,amsmath}
\usepackage{graphicx}
\usepackage{hhline}
\usepackage{textcomp}

\begin{document}

\raggedbottom

\title{High-resolution imaging of molecular collisions using a Zeeman decelerator}

\author{Vikram Plomp}
\affiliation{Radboud University Nijmegen, Institute for Molecules and Materials, Heijendaalseweg 135, 6525 AJ Nijmegen, the Netherlands}

\author{Zhi Gao}
\affiliation{Radboud University Nijmegen, Institute for Molecules and Materials, Heijendaalseweg 135, 6525 AJ Nijmegen, the Netherlands}

\author{Theo Cremers}
\affiliation{Radboud University Nijmegen, Institute for Molecules and Materials, Heijendaalseweg 135, 6525 AJ Nijmegen, the Netherlands}

\author{Matthieu Besemer}
\affiliation{Radboud University Nijmegen, Institute for Molecules and Materials, Heijendaalseweg 135, 6525 AJ Nijmegen, the Netherlands}

\author{Sebastiaan Y.T. van de Meerakker}
\affiliation{Radboud University Nijmegen, Institute for Molecules and Materials, Heijendaalseweg 135, 6525 AJ Nijmegen, the Netherlands}

\date{\today}

\begin{abstract}
We present the first crossed beam scattering experiment using a Zeeman decelerated molecular beam. The narrow velocity spreads of Zeeman decelerated NO ($X  ^2\Pi_{3/2}, j=3/2$) radicals result in high-resolution scattering images, thereby fully resolving quantum diffraction oscillations in the angular scattering distribution for inelastic NO-Ne collisions, and product-pair correlations in the radial scattering distribution for inelastic NO-O$_2$ collisions. These measurements demonstrate similar resolution and sensitivity as in experiments using Stark decelerators, opening up possibilities for controlled and low-energy scattering experiments using chemically relevant species such as H and O atoms, O$_2$ molecules or NH radicals. 
\end{abstract}

\maketitle
Establishing experimental tools to study molecular collisions with the highest possible level of detail has been an important goal in molecular physics for decades \cite{LevineBernstein1987}. The sensitivity and resolution of the experiment depends on the control over the particles before the collision, and how they are detected afterwards. In recent years, the combination of Stark deceleration and velocity map imaging (VMI) to both control and probe the quantum state and velocity of molecules, has greatly enhanced the possibilities to investigate molecular collisions in crossed beam experiments \cite{Onvlee2014}. The narrow velocity and angular spreads of Stark-decelerated beams result in scattering images with unprecedented radial and angular resolution, that can be exploited to resolve structure in the scattering images -- and thus the differential cross section (DCS) of the scattering process -- that would have been washed out using conventional molecular beams. Recent examples include the direct imaging of quantum diffraction oscillations \cite{VonZastrow2014,vogels2014high,onvlee2017imaging}, the measurement of correlated excitations in bimolecular collisions \cite{gao2018observation}, and the probing of scattering resonances at low collision energies \cite{vogels2015imaging,vogels2018scattering}. 

Despite these successes and further prospects to unravel fine details of collision processes, the Stark deceleration technique has a major limitation. As the method relies on the interaction of neutral molecules with electric fields, it can only be applied to species with a sufficiently large electric dipole moment. Although these include important molecules for scattering studies \cite{meerakker2012,Brouard2014}, many chemically relevant species like H, O and F atoms, O$_2$ molecules or ground state NH radicals exclusively have a magnetic dipole moment, rendering the Stark deceleration technique useless. Yet, these species are of paramount importance to molecular reaction dynamics \cite{Yang2007}, surface scattering \cite{Golibrzuch2015}, and the emerging fields of cold and ultracold molecules alike \cite{Bell2009}. 

Recently, various types of Zeeman decelerators -- the magnetic analogue of a Stark decelerator -- have been realized, and the successful deceleration \cite{Vanhaecke:PRA75:031402, Narevicius:NJP9:358, Narevicius:PRL100:093003, Lavert-Ofir:PCCP13:18948, Wiederkehr:JCP135:214202, Trimeche:EPJD65:263, Momose:PCCP15:1772, Dulitz:JCP140:104201, Cremers:PRA95:043415, Cremers:PRA98:033406, McArd2018} and subsequent trapping \cite{Hogan:PRL101:143001,Wiederkehr:PRA81:021402,Liu:PRA91:021403, Liu:PRL118:093201,Akerman:PRL119.073204,Segev2019} of a variety of atomic and molecular species has been reported. Yet, the application of molecular decelerators in crossed beam experiments poses specific requirements on density, state purity, and velocity control of the decelerated packets that are rather different from the requirements in trapping experiments, and therefore the prospects for using Zeeman decelerators in advanced crossed beam experiments have thus far remained elusive.

In this Letter, we report the first crossed beam experiments employing a Zeeman decelerator, using the inelastic scattering of NO radicals with Ne atoms and O$_2$ molecules as model systems for atom-molecule and molecule-molecule interactions. We use a 2.2-meter long Zeeman decelerator that has recently been constructed and that is specifically optimized for scattering experiments. State-to-state inelastic DCSs are probed using velocity map imaging. The narrow longitudinal and transverse velocity distributions of the packets of NO radicals emerging from the Zeeman decelerator result in scattering images with very high radial and angular resolutions. This is exploited to fully resolve narrowly-spaced quantum diffraction oscillations in the angular scattering distribution for NO-Ne collisions, and multiple close-lying rings pertaining to rotational product-pair correlations for NO-O$_2$ collisions. Resolving such structures has recently been a testimony of the unprecedented resolution afforded by Stark decelerators. The measurements presented in this Letter demonstrate that similar resolutions can now also be achieved using Zeeman decelerators, opening up the possibility to perform state-of-the-art cold and controlled collision experiments using a wide range of magnetic species, greatly enhancing the experimental possibilities and chemical diversity in this emerging field of research.

We chose the NO radical for these experiments, as this species has been frequently used in crossed beam experiments employing Stark decelerators, directly facilitating a comparison with the results from earlier work. A molecular beam of NO radicals with a mean velocity of 450 m/s is formed by expanding 5\% NO seeded in krypton into vacuum through a Nijmegen Pulsed Valve. Due to the supersonic expansion, the majority of the NO radicals will reside in the $j=1/2$ rotational level of the $v=0$ vibrational ground state of the $X\,^2\Pi_{1/2}$ electronic ground state. This level is split into two $\Lambda$-doublet components with $e$ and $f$ parity, that are close in energy and equally populated. 

The $^2\Pi_{1/2}$ spin-orbit manifold of the electronic ground state has negligible magnetic moment, as the orbital angular momentum and electron spin lead to equal but opposing contributions to the magnetic moment. Just after passing the skimmer, a part of the NO beam is therefore transferred to the $j=3/2$ level of the spin-orbit excited $^2\Pi_{3/2}$ manifold by inducing the $A\,^2\Sigma^+, N=0, j=1/2 \leftarrow X\,^2\Pi_{1/2}, v=0, j=1/2, f$ transition, followed by spontaneous decay to both spin-orbit manifolds of the $X\,^2\Pi$ electronic ground state.  As the Franck-Condon factors for the $A \rightarrow X$ transition are highly off-diagonal, only a small fraction of the excited NO radicals ends up in the $v=0$ vibrational ground state of the $X\,^2\Pi_{3/2}, j=3/2$ level \cite{Wang2013}. For NO-O$_2$ collisions, in which the scattered flux is distributed over various O$_2$ final states (\emph{vide infra}), we added a second dye laser to enhance population transfer into the $X\,^2\Pi_{3/2}, j=3/2$ initial level by stimulated emission pumping.

It is noted that the parity selection rules for the $A \leftrightarrow X$  transition in principle allow for the production of packets of NO radicals that are exclusively in the $X\,^2\Pi_{3/2}, j=3/2, e$ level. However, we find that both the $e$ and $f$ components of the $X\,^2\Pi_{3/2}, j=3/2$ level are approximately equally populated. This is assumed to be a result of the extremely small $\Lambda$-doublet splitting in this state in combination with the presence of stray electric fields inside the vacuum chamber.  

NO radicals in the $m_j=3/2$ component of the $X\,^2\Pi_{3/2}, j=3/2$ level have an effective magnetic moment of 1.2 $\mu_B$, and thus experience a significant Zeeman effect making them amenable to the Zeeman deceleration technique. Both the $e$ and $f$ $\Lambda$-doublet components have magnetic low-field seeking and high-field seeking $m_j$ levels, and NO radicals in both components are transported through the Zeeman decelerator with equal efficiency. The Zeeman decelerator used here consists of an alternating array of 100 solenoids and 100 hexapoles to control the longitudinal and transverse motions independently. A detailed description of the pulsed solenoids, hexapoles and electronics is given elsewhere \cite{Cremers2019}. Briefly, the solenoids are produced from a single winding of copper capillary through which cooling liquid is passed. Currents of up to 5 kA are pulsed through the solenoids using home-built electronics, producing a maximum magnetic field strength exceeding 2.5 Tesla inside the solenoid. The hexapoles consist of six arc-shaped permanent magnets with alternating magnetization direction. The Zeeman decelerator is operated at a repetition rate of 10 Hz. 
  
After exiting the Zeeman decelerator the packet of NO ($X\,^2\Pi_{3/2}, j=3/2$) radicals is guided towards the interaction region by 8 additional hexapoles, and scatters with a beam of Ne atoms or O$_2$ molecules that intercepts the Zeeman decelerator beam axis under an angle of 90$^{\circ}$ at a distance of 226 mm from the exit of the decelerator. The Ne and O$_2$ beams are produced using an Even Lavie valve held at 80 K and 120 K, resulting in a mean velocity of 465 m/s and 530 m/s, respectively. Scattered products are detected state-selectively using a recoil-free (1+1') resonance-enhanced multiphoton ionization scheme in combination with high-resolution velocity map imaging ion optics.

\begin{figure}[!htb]
   \centering
    \resizebox{1.0\linewidth}{!}
    {\includegraphics{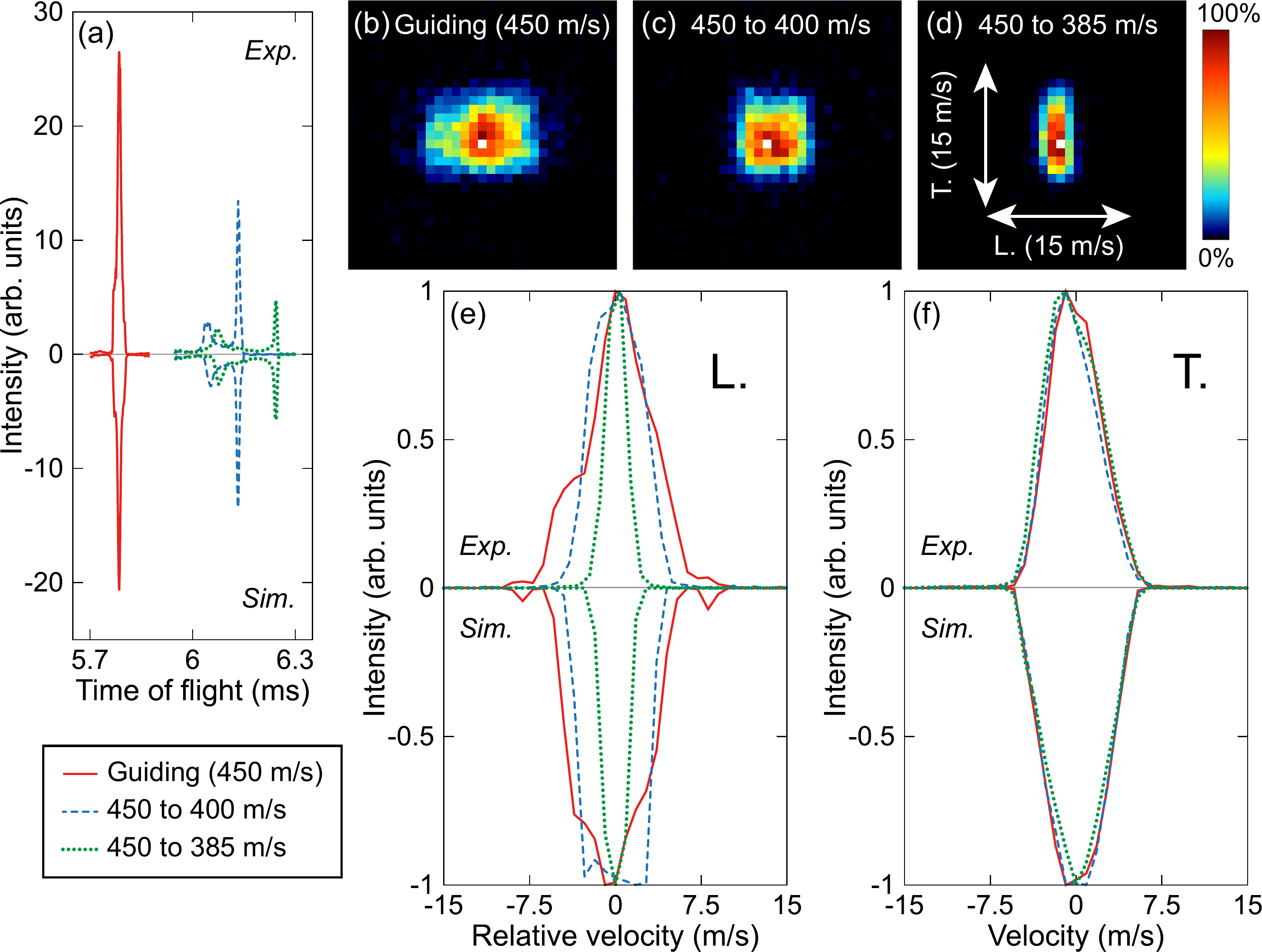}}
    \caption{(a) Selected parts of the TOF profiles for NO radicals ($X\,^2\Pi_{3/2}$) that exit the Zeeman decelerator when it is used to guide a packet of NO at a constant speed of 450 m/s, or decelerate it to a final velocity of either 400 m/s or 385 m/s. (b,c,d) The measured velocity-mapped ion images at the peak of the TOF profiles, together with the corresponding longitudinal (L.) and transverse (T.) velocity distributions of the packets displayed in panels (e) and (f), respectively. The experimental profiles (Exp.) are shown above the simulated profiles (Sim.).}
    %Measured velocity-mapped ion images of NO radicals ($X\,^2\Pi_{3/2}$) that exit the Zeeman decelerator when selecting an initial packet with a mean velocity of 450 m/s, and guiding it at constant speed (a) or decelerating it to a final velocity of 400 m/s (b) or 385 m/s (c). Corresponding longitudinal (d) and transverse (e) velocity distributions of the packets, where the experimental distributions are shown above the simulated distributions.}%The VMI resolution was calibrated to be 0.9 m/s per pixel.
    \label{fig:beamspots}
\end{figure}

The Zeeman decelerated packets of NO ($X\,^2\Pi_{3/2}, j=3/2$) radicals are characterized by recording both the time-of-flight (TOF) profiles of the decelerated radicals, and by measuring their velocity spreads using VMI. Figure \ref{fig:beamspots}a shows the TOF profiles that are measured when the decelerator is programmed to guide a packet of NO at a constant speed of 450 m/s, or decelerate it to a final velocity of either 400 m/s or 385 m/s. The velocity spread of the packets depends on the pulsing sequence of the Zeeman decelerator coils, and thus changes with the final velocity of the packet. This is illustrated by the VMI images shown in Fig. \ref{fig:beamspots}, which were measured at the peak of the TOF profiles and thereby reflect the velocity distributions of the most intense parts of the packets. Here, each pixel corresponds to a velocity of 0.9 m/s, as was carefully calibrated using a procedure originally developed for Stark decelerators \cite{Onvlee2014}. It is observed that the width of the longitudinal velocity distribution strongly depends on the final velocity of the packet, and ranges from 6.8 m/s (FWHM) for the highest velocity to 2.0 m/s for the lowest velocity (see Table \ref{StarkZeemanComparison}). By contrast, the transverse spread is independent of the final velocity, and amounts to about 5.5 m/s for all packets. This reflects the operation principles of our Zeeman decelerator: the decelerating forces are provided by the solenoids, and for ever higher deceleration rates, a packet of NO with ever smaller longitudinal velocity spread is produced. The transverse focusing forces are dominated by the hexapoles, and all molecules experience the same focusing forces independent of their forward velocity. Thus, the longitudinal deceleration and transverse focusing properties are effectively decoupled. These observations are well reproduced by the TOF profiles and velocity distributions resulting from numerical trajectory simulations shown in Fig. \ref{fig:beamspots}, which take into account both the $m_j=3/2$ and $m_j=1/2$ contribution.

As the resolution of a collision image critically depends on the velocity spreads of the decelerated beams, it is instructive to compare these spreads attained here for NO ($X\,^2\Pi_{3/2}$) using the Zeeman decelerator, with those routinely attained for NO ($X\,^2\Pi_{1/2}$) using a Stark decelerator. Table \ref{StarkZeemanComparison} summarizes these spreads, where we considered the Stark decelerator described in detail in Ref. \cite{Onvlee2014,Onvlee2014correction}. It is seen that the longitudinal velocity spreads for Zeeman-decelerated NO ($X\,^2\Pi_{3/2}$) can be made similar or even smaller compared to the longitudinal velocity spread of NO  ($X\,^2\Pi_{1/2}$) typically emerging from a Stark decelerator. By contrast, the transverse spread of the Zeeman-decelerated NO significantly exceeds the corresponding spread when using a Stark decelerator. This reflects the relatively strong focusing forces of the used magnetic hexapoles. It is noted that while the longitudinal and transverse spreads in the Stark-decelerated beams are coupled, these spreads can be tuned independently in the Zeeman decelerator. For instance, the transverse spread in the experiments reported here can simply be reduced by exchanging the hexapole magnets for weaker ones, without affecting the longitudinal properties of the packets.

\begin{table}[!h] 
\centering
\caption{Experimentally determined longitudinal (L.) and transverse (T.) velocity spreads of the packets of NO exiting our Zeeman decelerator, together with the spreads typically obtained when using a Stark decelerator as given in Ref. \cite{Onvlee2014correction}.}
\label {StarkZeemanComparison}
\begin{tabular}{lccc}
\hline
{} & Operation & \multicolumn{2}{c}{Velocity spread (FWHM)}\\
{} & mode & L. (m/s) & T. (m/s)\\
\hline
Stark \cite{Onvlee2014correction} $\text{ }$ & Guiding (480 m/s) & 4.8 & 1.4\\
Zeeman & Guiding (450 m/s) & 6.8 & 5.3\\
{} & 450 to 400 m/s & 5.2 & 5.2\\
{} & 450 to 385 m/s & 2.0 & 5.9\\
\hline
\end {tabular}
\end {table}

\begin{figure}[!htb]
   \centering
    \resizebox{1.0\linewidth}{!}
    {\includegraphics{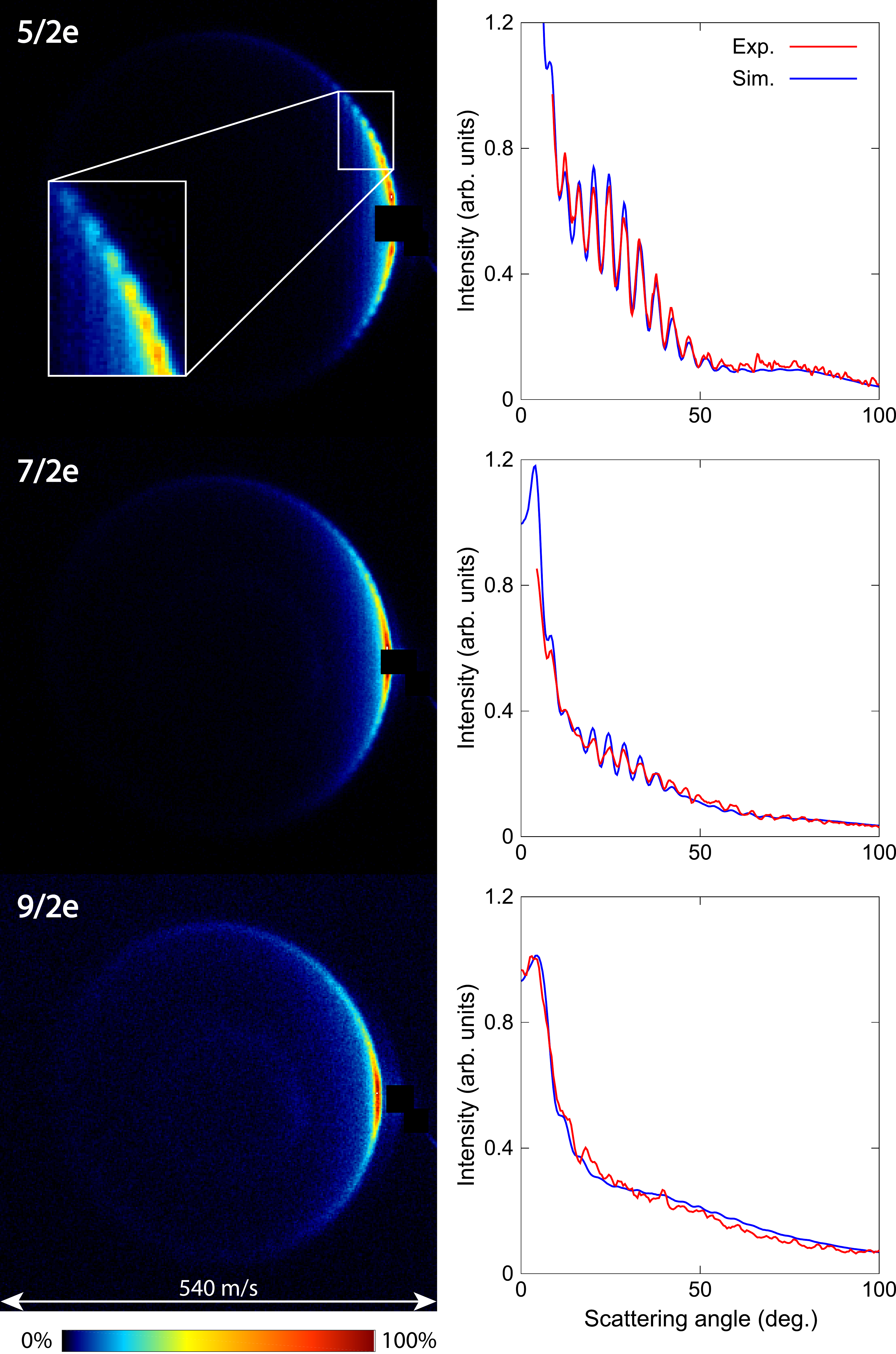}}
    \caption{Experimental scattering images (left panels) for the scattering processes NO ($X\,^2\Pi_{3/2}, j_{NO}=3/2$, e+f) + Ne $\rightarrow$ NO ($X\,^2\Pi_{3/2}, j'_{NO}$, e) + Ne. The corresponding experimental (Exp.) and simulated (Sim.) angular scattering distributions are shown in the right panels. The observed oscillatory structures originate from the diffraction of matter waves during the molecular collisions. For the $j'_{NO}=9/2$e state, a very weak inner ring is observed due to excitation from residual spin-orbit ground state NO ($X\,^2\Pi_{1/2}$) present in the molecular beam.}
    %For excitation into the $j=9/2$ state, a very weak inner ring is observed due to a overlapping REMPI transition that simulaneously probes the XXX state.}%The VMI resolution was calibrated to be 1.8 m/s per pixel.
    \label{fig:NO-Ne}
\end{figure}

\begin{figure}[!htb]
   \centering
    \resizebox{1.0\linewidth}{!}
    {\includegraphics{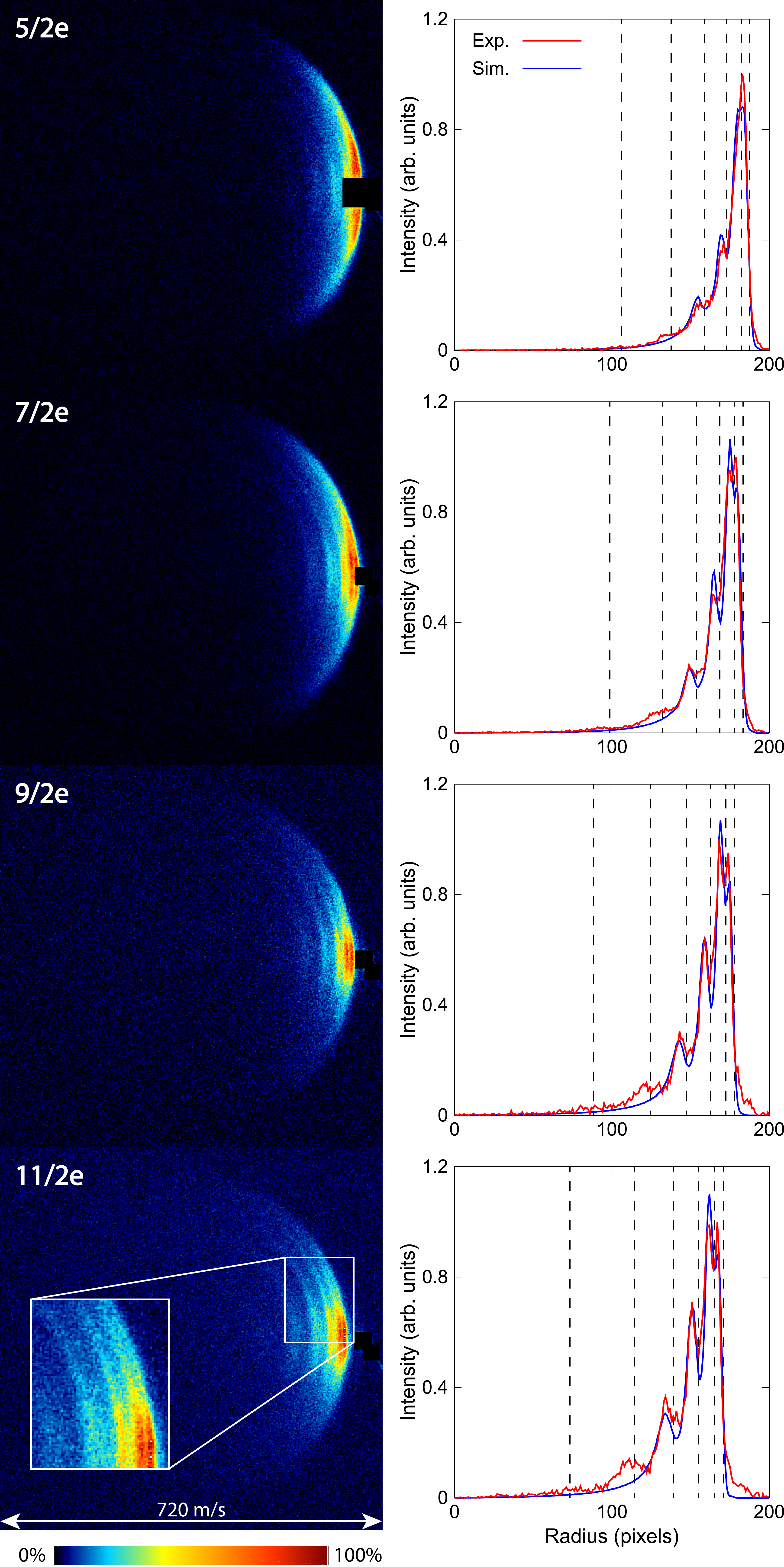}}
    \caption{Experimental scattering images (left panels) for the scattering processes NO ($X\,^2\Pi_{3/2}, j_{NO}=3/2$, e+f) + O$_2$ ($N_{O_2}=1$) $\rightarrow$ NO ($X\,^2\Pi_{3/2}, j'_{NO}$, e) + O$_2$ ($N'_{O_2}$). The corresponding experimental (Exp.) and simulated (Sim.) radial scattering distributions are shown in the right panels. The observed multiple concentric rings originate from correlated excitations in the NO radicals and O$_2$ molecules. The simulations are based on theoretical calculations that include only four O$_2$ states ($N'_{O_2}=1,3,5,7$). The vertical dashed lines indicate the kinematic cut-off positions, i.e. the expected radii for different final states of O$_2$.}%The VMI resolution was calibrated to be 1.8 m/s per pixel.
    \label{fig:NO-O2}
\end{figure}

The velocity spreads currently obtained with the Zeeman decelerator are sufficient to resolve fine structures in collision images. We demonstrate this by performing collision experiments with Ne and O$_2$, for which the packets of NO ($X\,^2\Pi_{3/2}, j=3/2$) radicals exiting the decelerator with a velocity of 385 and 400 m/s were chosen, resulting in collision energies of 183 and 285 cm$^{-1}$, respectively. The experimental scattering images for inelastic collisions of NO with Ne and O$_2$ that excite the NO radicals to various final states are shown in Figs. \ref{fig:NO-Ne} and \ref{fig:NO-O2}, respectively. Here, each pixel corresponds to a velocity of 1.8 m/s. The images are presented such that the relative velocity vector is directed horizontally, with forward scattering angles positioned at the right side of the image. Small segments of the images are masked around the forward direction, since the initial beam gives a contribution to the signal here.

For NO-Ne, clear oscillatory structures are observed, originating from the quantum mechanical nature of molecules that leads to diffraction of matter waves during the molecular collisions. The oscillations are most pronounced for excitation into the $j=5/2$ level, and have less contrast for excitation into higher rotational levels. For NO-O$_2$, multiple scattering rings are observed in a single image that originate from correlated excitations in both the NO and O$_2$ molecules. For a given image probing a specific transition in NO, each ring corresponds to a different ($ N_{O_2}=1 \rightarrow N'_{O_2}$) inelastic excitation in the O$_2$ molecule. The relative intensity of the inner rings is observed to increase as NO is excited to higher rotational states, consistent with earlier NO-O$_2$ work using Stark decelerators and with propensity rules for bimolecular collisions \cite{gao2018observation,gao2018correlated}.

To further illustrate the ability to resolve fine structures in the images, we analyzed the \emph{angular} and \emph{radial} scattering distributions for NO-Ne and NO-O$_2$, respectively (see Figs. \ref{fig:NO-Ne} and \ref{fig:NO-O2}). The angular scattering distributions are retrieved from the experimental image intensities within a narrow annulus around the rims of the rings, whereas radial scattering intensity distributions are determined within a narrow cone of the images at near-forward scattering angles. The experimentally obtained distributions can be directly compared to the distributions that result from simulated images based on \emph{ab initio} coupled-channels calculations that use state-of-the-art NO-Ne and NO-O$_2$ potential energy surfaces \cite{Cybulski2012,gao2018observation}. In these simulations, the theoretical state-to-state cross sections for the $e$ and $f$ initial state are taken into account with equal contribution. Our measurements are quantitatively reproduced by the simulated distributions. The spacing and phase of the diffraction structures, as well as the position and relative intensities of the pair-correlated rings are in excellent agreement with theoretical predictions. These results demonstrate that the high intrinsic resolution in experiments using Stark decelerators can now also be achieved in experiments using Zeeman decelerators.

In this Letter, we have reported the first crossed beam scattering experiment using the combination of Zeeman deceleration and velocity map imaging. Our direct observations of diffraction oscillations in NO-Ne collisions and product-pair correlations in NO-O$_2$ collisions illustrate that the image resolutions are sufficient to resolve narrow-spaced structures in both the angular and radial scattering distributions. Such structures have thus far only been observed using Stark decelerators, underlining the prospects for performing high-resolution and low-energy scattering experiments with magnetic particles using Zeeman decelerators. This yields the interesting outlook to use a variety of chemically relevant species in scattering experiments employing decelerators, that could not be used before. These include H, O ($^3P$), O ($^1D$) and F atoms, as well as molecular species such as O$_2$ and NH. These species have been of paramount importance in benchmark reactive scattering studies in the past, and are prime candidates to study barrier-less reactions at low energies. Cold and controlled crossed beam experiments using a Zeeman decelerator as reported here offer new and exciting prospects to study collisions and reactions with unprecedented level of precision and in unexplored energy regimes.       

\section{acknowledgments}
This work is part of the research program of the Netherlands Organization for Scientific Research (NWO). S.Y.T.v.d.M. acknowledges support from the European Research Council (ERC) under the European Union's Seventh Framework Program (FP7/2007-2013/ERC Grant Agreement No. 335646 MOLBIL) and from the ERC under the European Union's Horizon 2020 Research and Innovation Program (Grant Agreement No. 817947 FICOMOL). We thank Gerrit Groenenboom and Ad van der Avoird for fruitful discussions about parity mixing and scattering calculations. We thank Niek Janssen, Edwin Sweers, Andr\'e van Roij and Michiel Balster for expert technical support.

\bibliographystyle{apsrev}
\bibliography{2019_NOZeeman,zeeman}

\end{document}